\documentclass[aip,pop,amsmath,amssymb,reprint]{revtex4-1}
\usepackage[T1]{fontenc} 
\usepackage{graphicx}
\usepackage{dcolumn}
\usepackage{bm}
\usepackage{comment}
\usepackage{soul}
\usepackage[dvipsnames]{xcolor}
\usepackage{color}

\begin{document}

\preprint{}

\title{Extreme high field plasmonics: electron acceleration and XUV harmonic generation from ultrashort surface plasmons}

\author{A. Macchi}\thanks{Invited speaker.}
\affiliation{National Institute of Optics, National Research Council (CNR/INO),  Adriano Gozzini laboratory, via Giuseppe Moruzzi 1, 56124 Pisa, Italy}\email{andrea.macchi@ino.cnr.it}
\affiliation{Enrico Fermi Department of Physics, University of Pisa, largo Bruno Pontecorvo 3, 56127 Pisa, Italy}

\author{G. Cantono}
\affiliation{Department of Physics, Lund University, PO Box 118, SE-22100, Lund, Sweden}

\author{L. Fedeli}\thanks{Presently at CEA/DSM/IRAMIS/LIDYL, 91191 Gif-sur-Yvette, France}
\affiliation{Department of Energy, Politecnico di Milano, 20133 Milano, Italy}

\author{F. Pisani}
\affiliation{Enrico Fermi Department of Physics, University of Pisa, largo Bruno Pontecorvo 3, 56127 Pisa, Italy}

\author{T. Ceccotti}
\affiliation{CEA/DSM/IRAMIS/LIDYL, CEA Saclay, 91191 Gif-sur-Yvette, France}

\date{\today}

\begin{abstract}
Experiments on the excitation of propagating surface plasmons (SPs) by ultrashort, high intensity laser interaction with grating targets are reviewed. At intensities exceeding $10^{19}~\mbox{W cm}^{-2}$ on target, i.e. in the strongly relativistic regime of electron dynamics, multi-MeV electrons are accelerated by the SP field as dense bunches collimated in a near-tangent direction. By the use of a suitable blazed grating, the bunch charge can be increased up to $\approx $660 picoCoulomb. Intense XUV high harmonics (HHs) diffracted by the grating are observed when a plasma with sub-micrometer scale is produced at the target surface by a controlled prepulse. When the SP is excited, the HHs are strongly enhanced in a direction quasi-parallel to the electrons. Simulations suggest that the HHs are boosted by nanobunching in the SP field of the electrons which scatter the laser field. Besides the static and dynamic tailoring of the target density profile, further control of electron and HH emission might be achieved by changing the SP duration using a laser pulse with rotating wavefront. This latter technique may allow to produce nearly single-cycle SPs. 
\end{abstract}

\maketitle

\section{Introduction}

Plasmonics is the study and exploitation of collective electron excitations, commonly referred to as plasmons, in solid density materials. Most plasmonics is actually based on surface modes, which exist at the boundary between two different media (including the case in which one medium is the vacuum), and can be both localized and propagating, the latter being known as surface plasmon polaritons (SPPs). Central to plasmonics is the coupling between surface plasmons and electromagnetic (EM) waves in vacuum, i.e. laser light irradiating the material.

Despite the apparent common ground between plasmonics and plasma physics, the interaction between the two research communities has been limited, as wisely recognized by G.~Manfredi in his Editorial\cite{manfrediPoP18} opening the Special Topic issue on ``Plasmonics and solid state plasmas'' in this journal. When considering plasmonics as part of solid state physics one argues that it has to deal with the quantum theory of condensed matter, while for most of ``traditional'' plasma physics a classical approach is adequate because of the high energy density. So viewing plasmonics and plasma physics as well separated areas might be considered as natural. However, the key concepts of surface and bulk plasmons and of their coupling with light can be described and modeled within classical electrodynamics. Hence, there is a potential overlap between plasmonics and laser-plasma interactions -- to be more specific, the research area of the interaction of high-intensity, femtosecond laser pulses with solid targets. In fact, while the short duration of the pulse preserves the high density and the sharp target-vacuum interface (both necessary to the surface plasmon excitation), the laser field strength allows to create free electrons by instantaneous field ionization, so that any material behaves similarly to a simple metal, i.e. a collisionless plasma, and can sustain surface plasmons over a broad range of frequencies.

One may also argue that plasmonics is oriented towards the development of advanced applications, such as sensors and optical devices, which require a level of control and reliability on the system that is far from the typical conditions of laser-plasma interaction experiments. It is true that a classical high-temperature plasma is somewhat a ``wild'' enviroment which is affected by various instabilities and nonlinear effects making its dynamics difficult to control, but this very general issue did not prevent the development of many successful plasma-based technologies. Moreover, the control and reproducibility of intense laser-plasma interaction phenomena has often been limited to imperfections of high power laser systems, most of which have become less severe thanks to recent developments. A good example, of direct relevance for the contents of the present paper, is provided by early attempts of exciting surface plasma waves (i.e. SPPs in a simple metal, from now on referred to simply as surface plasmons or SPs) by irradiating solid targets with intense femtosecond pulses\cite{gauthierSPIE95}. In this regime, the linear coupling of SPs with laser light requires a target with a periodical surface modulation, i.e. a grating. High power femtosecond pulses, amplified via the Nobel prize--awarded Chirped Pulse Amplification technique\cite{stricklandOC85}, are typically accompanied by both short prepulses and long pedestals preceding the main short pulse: these spurious emissions may be intense enough to damage and pre-ionize the target, causing early plasma formation and pre-expansion able to destroy a shallow modulation at the surface. This issue has limited experiments on high-intensity interaction with structured targets, and thus on high field SP excitation\cite{kahalyPRL08,huPoP10,bagchiPoP12} to moderate intensities for a long time. Recently, the development of ``pulse cleaning'' techniques such as the ionization shutter or plasma mirror\cite{kapteynOL91,dromeyRSI04,levyOL07,thauryNP07} delivered pulses with extremely high values of pulse-to-prepulse intensity ratio, briefly named as contrast. This achievement makes now possible to investigate the interaction of the most intense pulses available today with targets having sub-micrometer thickness and structuring. 

In a series of experiments performed with the UHI100 laser at the SLIC facility of CEA Saclay in France, our group has investigated the interaction of high intensity ($I>10^{19}~{\rm W cm}^{-2}$), short duration ($\tau=25$~fs), ultrahigh contrast pulses with grating targets. 
Noticeably, the laser frequency and intensity are such that the electron dynamics is strongly relativistic. In these conditions, the properties of SPs are not well known. As a matter of fact, all the assumptions underlying the simple linear theory of SPs break down\cite{macchiPoP18}.
However, these experiments provided strong evidence of SP-enhanced emission of energetic particles such as protons\cite{ceccottiPRL13}, electrons\cite{fedeliPRL16} and photons as extreme ultraviolet (XUV) harmonics of the incident laser\cite{cantonoPRL18}.  
Hence, these observations may stimulate theoretical advances in nonlinear plasmonics as well.

In the most recent experiments, we showed that features of electron and XUV harmonic emissions can be enhanced by, respectively, shaping the grating profile\cite{cantonoPoP18} and smoothing the density gradient using a femtosecond prepulse\cite{cantonoPRL18}. Both cases confirm that the high-contrast interaction is sensitive to the target density profile at a sub-micrometer scale and show the possibility of improved control of the latter. 

The results on XUV emission, supported by detailed simulations\cite{cantonoPRL18}, enlighted a new mechanism of harmonic generation which is strongly correlated with electron acceleration by the SP. The harmonic emission is enhanced by nanobunching of electrons trapped in the SP field, which is somewhat reminiscent of the collective beam instability in a free electron laser\cite{pellegriniRMP16}. The mechanism thus represents a novel example of exploiting self-organization and nonlinear dynamics to develop a plasma-based device or technology.

Secondary emissions from femtosecond laser-irradiated solid targets are characterized by the ultrashort duration, which is of the order of the laser pulse duration at the source. For specific applications it is desirable to push the source duration down to the attosecond regime, a goal partly reached in the case of XUV harmonics which constitute a train of sub-femtosecond spikes. Inspired by a scheme for the spatial separation of the spikes\cite{wheelerN12,vincentiPRL12}, we proposed a scheme for the excitation of extremely short SPs with duration approaching the single-cycle limit\cite{pisaniACSP18}. The concept exploits the wavefront rotation (WFR) of a short laser pulse\cite{quereJPB14}, which is another possible control parameter of the interaction. Near-single-cycle SPs might be of interest for applications in both ordinary and high-field plasmonics\cite{macdonaldNP09,dombiOE10,krugerN11,raczAPL11,liPRL13,polyakovPRL13}. 

In the remainder of the paper, after briefly recalling the basics of SP excitation by laser coupling with a grating we review the above mentioned results on electron acceleration and XUV harmonic generation. We also include a discussion of the simple modeling of electron acceleration in the SP field and of SP generation using a pulse with WFR. Other SP-driven effects such as enhanced heating and proton acceleration or the analysis of SP excitation in a grating target are discussed in detail in previous papers\cite{ceccottiPRL13,sgattoniPPCF16,macchiPoP18} and references therein.

\section{Surface plasmon excitation in gratings}

In this section we summarize the basics of SP excitation by a laser pulse impinging on a grating, commenting on specific issues of the high field regime. 
We consider the interface between vacuum and a cold plasma or simple metal with dielectric function $\varepsilon(\omega)=1-\omega_p^2/\omega^2$, where $\omega_p=(4\pi e^2n_e/m_e)^{1/2}$ is the plasma frequency and $n_e$ the electron density. The dispersion relation of SPs is
\begin{equation}
k_{\rm SP}=\frac{\omega}{c}\left(\frac{\omega^2_p/\omega^2-1}{\omega^2_p/\omega^2-2}\right)^{1/2} \; ,
\label{eq:SPdisp}
\end{equation}
provided that $\omega<\omega_p/\sqrt{2}$. Eq.(\ref{eq:SPdisp}) prevents phase matching of the SP with an obliquely incident EM plane wave because the latter has a wavevector component parallel the surface equal to $k_{\parallel}=(\omega/c)\sin\phi$, with $\phi$ the angle of incidence, thus $k_{\parallel}<k_{\rm SP}$. However, if the surface is modulated with period $d$, the SP dispersion relation is replicated long the $k$-axis with period $\pi/d$ or, equivalently, folded into the Brillouin zone $|k|<\pi/d$ (Floquet--Bloch theorem), so that intersections with the EM dispersion relation are now possible. This grating coupling scheme is not the only one used in ordinary plasmonics \cite{bookMaier2007} but it is the one that appears most suitable when exciting SPs with intense laser pulses. In fact, when using prism-based schemes (e.g. Kretschmann or Otto configurations) the intense laser pulse would propagate through layers of dielectric material, thus undergoing distortions by nonlinear effects and ionization. Given the grating period $d$, the resonant excitation of SPs occurs at incidence angles $\phi_{\rm res}$ such that
\begin{equation}
\sin\phi_{\rm res}=\pm\frac{c}{\omega}k_{\rm SP}+n\frac{\lambda}{d} \; ,
\label{eq:SPres}
\end{equation}
where $\lambda=2\pi c/\omega$ is the wavelength in vacuum and $n=\pm 1,\pm 2, \ldots$ is an integer. Notice that the SP may also propagate in the direction opposite to the impinging pulse.

Eq.(\ref{eq:SPres}) assumes that the dispersion relation (\ref{eq:SPdisp}) is not strongly modified by the surface modulation, which is a good approximation for shallow gratings since deviations are of the order $(\delta/d)^2$ with $\delta$ the grating depth. In the case of our experiments the gratings are not extremely shallow, since $d \sim \lambda$ and $\delta=(0.3\--0.5)\lambda$, a choice made to preserve the grating against hydrodynamical expansion. 
At this stage, however, it is not important to take such corrections into account since Eq.(\ref{eq:SPdisp}) may be already considered as an approximation in the high field regime, because nonlinear and kinetic effects have also been neglected.

In a solid density material and for optical frequencies $\omega_p\gg\omega$, so that Eq.(\ref{eq:SPres}) may be simplified as $\sin\phi_{\rm res}=\pm 1+n{\lambda}/{d}$, which is equivalent to the condition that the $n$-th diffraction order from the grating is along the surface. 
This observation can be useful when one considers that in experiments with superintense, tightly focused laser pulses, only a few periods of the grating are illuminated, and the assumption of an infinite periodic medium becomes questionable. However, one may also think to the SPs as excited by the laser light scattered by the grating in conditions of constructive interference, and argue that the finite spot size may affect the resonance width.

\section{Electron acceleration by surface plasmons}

In the low field regime, plasmonic-enhanced emission of photoelectrons has been widely studied with regard to the development of ultrafast photocathodes\cite{tsangPRB91,dombiOE08,hwangPRB09,raczAPL11,watanabeJAP11,liPRL13,gongPRApp14,onoJPD15}. In the case of laser-grating interactions, some experiments found photoelectrons of anomalously high energy, attributed to ponderomotive acceleration in the evanescent SP field\cite{zawadzkaAPL01,kupersztychPRL01,kupersztychPoP04,irvinePRL04,irvinePRA06}. The features of electron acceleration emerging in the case of high field, relativistically strong interactions are qualitatively different from such previous observations.

\subsection{Experimental observation}

\begin{figure}[tb]
\includegraphics[width=0.48\textwidth]{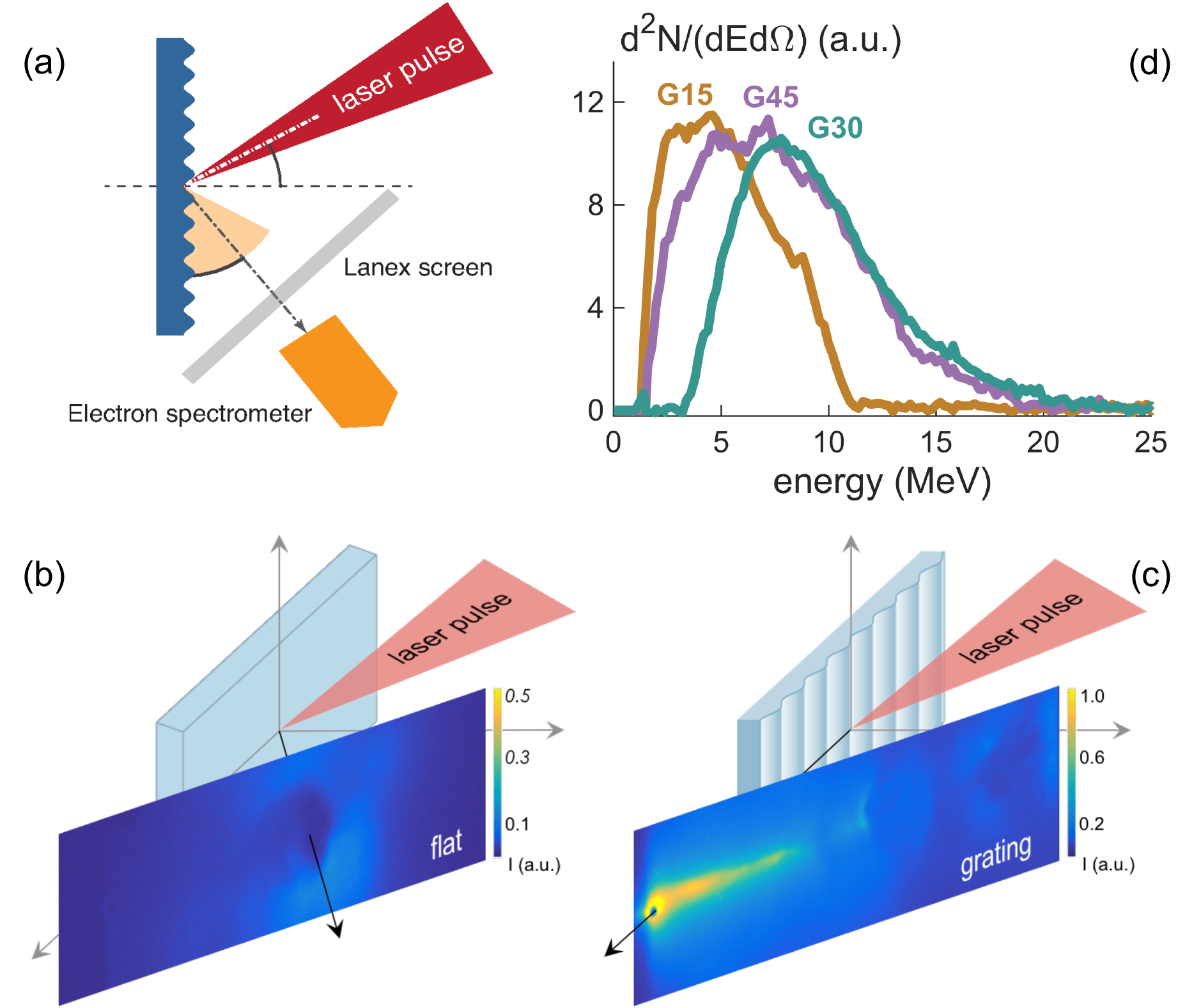}
\caption{Electron acceleration by surface plasmons\cite{fedeliPRL16,cantonoPoP18}. a): the basic experimental set-up (see text for laser parameters).
b): the image on the lanex screen for a flat target, showing a diffuse angular distribution of electrons. c): the image for a grating target irradiated at the resonant angle for SP excitation ($30^{\circ}$ for the case shown), showing an highly collimated emission close to the tangent at the target surface. d): energy spectrum for the collimated electron emission, for three gratings having different resonant angles ($15^{\circ}$, $30^{\circ}$, $45^{\circ}$) corresponding to grating periods $d=1.35\lambda$, $2\lambda$ and $3.4\lambda$, respectively, where $\lambda=0.8~\mu$m. The gratings had a sinusoidal profile with a depth of $0.4\lambda$.
Adapted and partially reproduced from G.~Cantono et al, Physics of Plasmas \textbf{25}, 031907 (2018), with the permission of AIP Publishing.
\label{fig:electrons}}
\end{figure}

The most striking effect observed in the experiments is the acceleration of collimated bunches of multi-MeV electrons on the vacuum side in a near tangent direction when the gratings are irradiated at the angle of incidence for which excitation of a SP is expected\cite{fedeliPRL16}. Fig.\ref{fig:electrons} summarizes the basic set-up and the main results of devoted experiments at SLIC (details can be found in Refs.\onlinecite{fedeliPRL16,cantonoPoP18}). The 25~fs duration, 0.8~$\mu$m wavelength laser pulse of the UHI-100 laser was focused by a $f/3.75$ parabola on gratings of different spatial period $d$. The pulse-to-prepulse contrast was $10^{12}$ and $10^{10}$ at, respectively, 20 and 5~ps before the main pulse. 

For all the gratings, the electron spectra near the resonant angle are broad but strongly non-thermal, peaking at energies of $\sim 8$~MeV and extending up to the highest detectable value of $\sim 18$~MeV for laser intensities in the range of $I=1.7\--3.4 \times 10^{19}~\mbox{W cm}^{-2}$. This corresponds to a dimensionless intensity parameter
\begin{equation}
a_0=0.85\left(\frac{I\lambda^2}{10^{18}~\mbox{W cm}^{-2}}\right)^{1/2}=2.8\--4.0 \; ,
\end{equation}
which indicates that the interaction is in the relativistic regime ($a_0>1$). 
The electrons are emitted at grazing incidence, typically within an angle of $\sim 10^{\circ}$ from the target tangent. Detailed measurements show that the electrons are collimated in a cone of $\sim 5^{\circ}$ aperture for all gratings, and that the charge flux exceeds nC/sr values. By varying the incidence angle of $5^{\circ}$ around the resonant value, the peak energy, charge and collimation of the electrons decrease by at least a factor of $\sim 2$.

The above findings are in sharp contrast with the electron emission from flat targets, which is spatially diffuse and limited to sub-MeV energies, below the detection threshold of the spectrometer. Such low energies are consistent with the estimate based on the so-called ponderomotive energy $U_p$ in the investigated energy range,  
\begin{equation}
U_p=m_ec^2\left((1+a_0^2/2)^{1/2}-1\right)=0.6-1~\mbox{MeV} \; .
\end{equation} 

\subsection{Simple model}

A simple model highlights the mechanism of electron acceleration by SP and provides an estimate for the attainable energy and the angle of emission. The model is similar to that used in the famous paper by Tajima and Dawson\cite{tajimaPRL79} on electron acceleration in wake plasma waves, the main difference being the essential two-dimensional (2D) nature of the SP. 

We consider a SP propagating in the laboratory frame $S$ along the surface of a step-boundary plasma with the density profile $n_e=n_0\Theta(-x)$ (i.e. the vacuum region is $x>0$). 
Assuming the SP to propagate along $y$ with phase velocity $v_p=\omega/k<c$ as given by the dispersion relation (\ref{eq:SPdisp}) (in this section  we write $k$ for $k_{\rm SP}$ since no confusion is possible), the SP fields have the form
$f(x,y,t)=\mbox{Re}\left(\tilde{f}(x){\rm e}^{iky-i\omega t}\right)$ 
with 
\begin{eqnarray}
\tilde{E}_y(x)&=&E_{\mbox{\tiny SP}}\left(\Theta(+x){\rm e}^{-q_{<}x}+\Theta(-x){\rm e}^{+q_{>}x}\right)  \;, \\
\tilde{B}_z(x) 
&=& \frac{i\omega /c}{q_{<}}
E_{\mbox{\tiny SP}}\left(\Theta(+x){\rm e}^{-q_{<}x}
+\Theta(-x){\rm e}^{+q_{>}x}\right) \; ,
\\
\tilde{E}_x(x) 
&=& -ik E_{\mbox{\tiny SP}} \left(\Theta(+x)\frac{{\rm e}^{-q_{<}x}}{q_{<}}
                              -\Theta(-x)\frac{{\rm e}^{+q_{>}x}}{q_{>}} \right) ,
\label{eq:SW_fields}
\end{eqnarray}
where $q_{>}=(\omega^2_p/\omega^2-1)^{1/2}k$ and $q_{<}=(\omega^2_p/\omega^2-1)^{-1/2}k$, and $E_{\mbox{\tiny SP}}$ is the amplitude of the longitudinal electric field($E_y$) component at $x=0$.
Notice that $E_y$ and $B_z$ are continuous at $x=0$ while $E_x$ is discontinuous, which implies a surface charge density $\sigma(y,t)=\mbox{Re}\left(\tilde{\sigma}{\rm e}^{iky-i\omega t}\right)$. 

We now perform a Lorentz transformation to the $S'$ frame moving with $v_p=\beta c$ along $y$ (so that $\gamma=(1-\beta^2)^{-1/2}=(\omega_p^2/\omega^2-1)^{1/2}$).  In this frame, the phase $ky-\omega t=k'y'$ with $k'=k/\gamma$ and $\omega'=0$, and the fields do not depend on time. The field amplitudes transform according to $E'_x=\gamma(E_x+\beta B_z)$, $E'_y=E_y$, $B'_z=\gamma(B_z+\beta E_x)$, 
and thus we obtain 
\begin{eqnarray}
\tilde{E}'_y(x)&=&E_{\mbox{\tiny SP}}\left(\Theta(+x){\rm e}^{-q_{<}x} +\Theta(-x){\rm e}^{+q_{>}x}\right) \;  ,\\
\tilde{B}'_z(x)&=&
\frac{i\omega^2_p}{kc\omega}E_{\mbox{\tiny SP}}\Theta(-x){{\rm e}^{+q_{>}x}} \; , \\
\tilde{E}'_x(x)&=&-\frac{i\gamma k}{q_{<}}E_{\mbox{\tiny SP}}
     \left(\Theta(+x){{\rm e}^{-q_{<}x}}\frac{1}{\omega^2_p/\omega^2-1}\right. \nonumber \\
     & & \left. -\Theta(-x){{\rm e}^{+q_{>}x}} \right) \; . 
\end{eqnarray}
Notice that $q_{<}={k}/{\gamma}=k'$ and
that the magnetic field $B'_z$ in $S'$ vanishes for $x<0$ and is discontinuous at the surface. This is consistent with the transformation of the surface charge density $\sigma$ in $S$, which yields a  surface current $\iota'=-\mbox{Re}\left(\beta\gamma c \sigma{\rm e}^{ik'y'}\right)$ in $S'$.

The electric field components can be derived from the electrostatic potential
$\Phi'=\Phi'(x,y)=\mbox{Re}\left(\tilde{\Phi}'(x){\rm e}^{ik'y'}\right)$ where
\begin{eqnarray}
\tilde{\Phi}'=
\frac{i\gamma}{k}E_{\mbox{\tiny SP}}
                 \left(\Theta(+x){\rm e}^{-q_{<}x} +\Theta(-x){\rm e}^{+q_{>}x}\right)\; .  
\end{eqnarray}
Since in the vacuum region ($x>0$) we have $B'_z=0$ in $S'$, the electron motion in this region can be simply described as the downhill motion in the 2D potential energy hill $U(x,y)=-e\Phi'(x,y)$.
The energy gain of a test electron moving in the such potential depends on the initial conditions, i.e. on the injection mechanism in the SP field. 
The most favorable case is that of an electron with initial velocity equal to the phase velocity ${\bf v}_p=v_p \hat{\bf y}$ (i.e. at rest in the moving frame $S'$) and initially placed at $x=0$ and $y'=\pi/2k'$ (modulus $2\pi/k'$). When such electron receives an infinitesimal kick in the $y$ direction and fully descends the potential hill along the $x=0$ plane, its energy gain is $W'=W_{\rm max}\equiv 2eE_{\rm SP}\gamma/k$ which corresponds to the energy-momentum four-vector
\begin{eqnarray}
(U'_f,{\bf p}'c)=(m_ec^2+W_{\rm max},0,p'_{fy},0) \; ,
\end{eqnarray}
with $U_f^{'2}=p^{'2}_{fy}c^2+m_e^2c^4$. Now, assuming $v_p\simeq c$ and since $\gamma \sim \omega_p/\omega \gg 1$, we can write $W_{\rm max} \simeq 2m_ec^2a_{\rm SP}\omega_p/\omega$ where $a_{\rm SP}=(eE_{\rm SP}/m_e\omega c)$. For a high amplitude SP with $a_{\rm SP} \sim 1$, we obtain $W_{\rm max}\gg m_e c^2$ and thus $U'_f\simeq W' \simeq p'_{fy}c$. Transforming back to the laboratory frame we obtain for the electron energy
\begin{eqnarray}
U_f=\gamma(U'_f+\beta p'_{fy}c) \simeq 2W'\frac{\omega_p}{\omega}=4m_ec^2a_{SP}\frac{n_0}{n_c} \; ,
\label{eq:maxgain}
\end{eqnarray}
where $n_c=m_e\omega^2/4\pi e^2$ is the cut-off density for EM waves. (As a function of the wavelength $\lambda=2\pi c/\omega$, $n_c=1.1 \times 10^{21}~\mbox{cm}^{-3}(\lambda/1~\mu\mbox{m}){-2}$.) 
It is curious to notice that expression (\ref{eq:maxgain}) corresponds to the maximum energy gain in a 1D sinusoidal wakefield\cite{tajimaPRL79} but for $n_0/n_c$ replaced by its inverse. 
However, in the SP field any small kick in the $-x$ direction will make the electron descend the potential hill on the vacuum side. Moreover, the steady magnetic field $B'_z$ for $x>0$ also bends the trajectory of an electron with $v'_y>0$ towards the vacuum region. Hence, most electrons acquire a finite $v'_x<0$ and a final energy $U'_f<W_{\rm max}$ in $S'$. As a representative case we consider an electron with same initial conditions as before, but with an initial infinitesimal kick in the $-x$ direction. When such electron eventually leaves the SP field region, it has acquired an energy $W'=W_{\rm max}/2$ and its energy-momentum is
\begin{eqnarray}
(U'_f,{\bf p}'c)=(m_ec^2+W_{\rm max}/2,p'_{fx},0,0) \; ,
\end{eqnarray}
with $U_f^{'2}=p^{'2}_{fx}c^2+m_e^2c^4$. Proceeding as above we obtain in $S$
\begin{eqnarray}
(U_f,{\bf p}_fc) \simeq (\gamma U'_f,U'_f,\gamma U'_f,0) \; ,
\end{eqnarray} 
so that the electron emerges with an energy $U_f$ and 
at an angle $\phi_e$ (with respect to the normal) given by
\begin{eqnarray}
U_f \simeq  m_ec^2a_{SP}\frac{n_0}{n_c} \; , \qquad \tan\phi_e=\frac{p_y}{p_x}\simeq\gamma \; .
\label{eq:energy-angle}
\end{eqnarray} 
The picture emerging from the simple model is in qualitative agreement with the experimental observations. For instance, the values of $8~MeV$ and $18~MeV$ observed for the peak and the highest energies at a resonant angle of $30^{\circ}$ corresponds to $\gamma \simeq 17$ and $37$ yielding $\phi_e=86.6^{\circ}$ and $88.5^{\circ}$, respectively, which is quite close to the value obtained from the electron beam imaging. 
A quantitative comparison with the formula (\ref{eq:energy-angle}) for the energy $U_f$ requires to estimate the amplitude $a_{\rm SP}$ of the SP and the exact electron density, which is possible only in simulations. The three-dimensional particle-in-cell (PIC) simulations reported in Ref.\onlinecite{fedeliPRL16} with $n_0=50n_c$ and other parameters matching the experimental ones show that $a_{\rm SP} \simeq 1$ and the electron energy extends up to $\sim 20$~MeV, which is fairly consistent with the model prediction of $U_f \simeq 25$~MeV; the shape of the energy spectrum was also reproduced. 

It may be argued that the simple model predicts much higher energies than those currently observed, if considering the possibilities of strongly relativistic SPs ($a_{\rm SP}\gg 1$), density values corresponding to full ionization of a solid target ($n_0>10^2n_c$)  and ``luckiest'' injection phases (see the discussion below) so that the limit (\ref{eq:maxgain}) is approached. While all these possibilities deserve further theoretical and experimental investigation, it is worth noticing that the acceleration length required to achieve the final energy in Eq.(\ref{eq:energy-angle}) is  
\begin{eqnarray}
L=\frac{U_f}{eE_{\mbox{\tiny SP}}}
\simeq \frac{\lambda}{2\pi}\frac{n_0}{n_c} \; ,
\end{eqnarray}
and four times larger when taking (\ref{eq:maxgain}) for $U_f$: the resulting values may exceed the typical diameter of the laser spot. This may be a limiting factor for the energy gain if the SP needs to be sustained by the laser field against strong damping or, simply, if the SP cannot propagate outside of the laser spot where the target material is ionized. Focusing the laser energy in an elongated spot (line focus) could overcome such limitation. 

It has also to be stressed that the injection of electrons in the SP field with an optimal phase is a critical factor to reach the maximum energy gain. The experimental results, with large amounts of electrons being accelerated to energies of the order of $U_f$, suggest that self-injection is quite efficient (although it delivers a broad spectrum). This is also supported by the study of Riconda et al.\cite{ricondaPoP15} based on the numerical solution of test particle motion in the fields of a high-amplitude SP. In particular, injection is provided to a large extent (in the laboratory frame) by the nonlinear magnetic force $F_y=-ev_xB_z/c$ along the SP propagation direction. An advanced study of injection should take into account the nonlinear modification of the SP field for high amplitudes, in the relativistic regime and close to the wavebreaking limit which should characterize any longitudinal wave: these are also issues for future theoretical modeling.

\subsection{Source optimization with blazed gratings}

\begin{figure}[tb]
\includegraphics[width=0.48\textwidth]{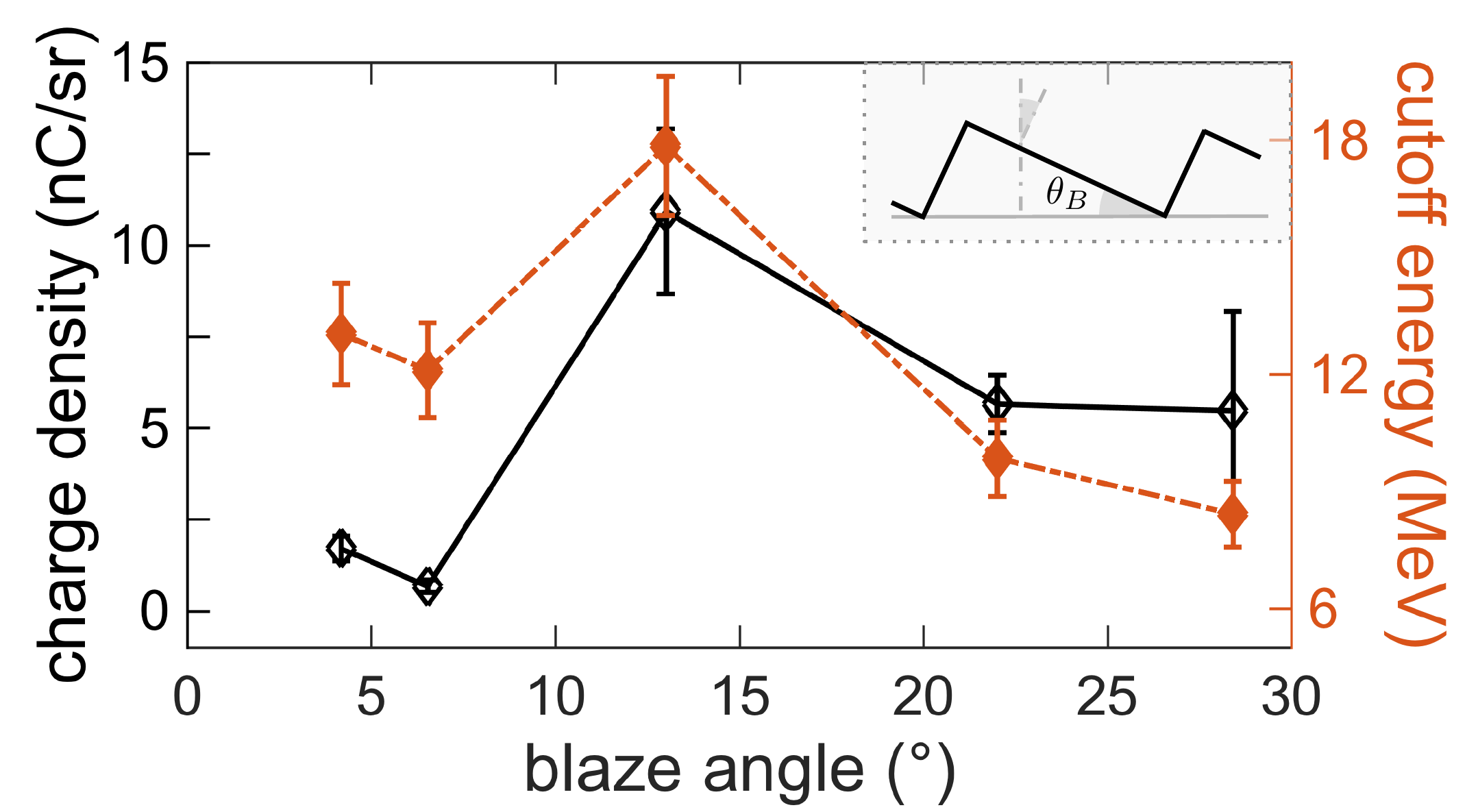}
\caption{Optimization of electron acceleration by using blazed gratings\cite{cantonoPoP18}. The total charge of the electron bunches and the cut-off energy in the detected spectrum are shown as a function of the blaze angle $\theta_B$ (see inset for the definition). The resonant angle was $30^{\circ}$ for the grating used, corresponding to $d=2\lambda=1.6~\mu$m. 
Adapted and partially reproduced from G.~Cantono et al, Physics of Plasmas \textbf{25}, 031907 (2018), with the permission of AIP Publishing.
\label{fig:blazed}
}
\end{figure}

Blazed gratings (BGs), having an asymmetrical triangular profile, are typically used to maximize the intensity of a particular order of diffraction in the reflection from a grating. To this aim, the blaze angle is chosen in order that the incident light is locally reflected in the same direction as the diffraction angle. In the high intensity regime, blazed gratings have been used to optimize high harmonic generation\cite{yeungNJP13,zhangOE17} (section~\ref{sec:harmonics}).

Following this principle, BGs may be used for more efficient excitation of a SP choosing the blaze angle so that the maximum efficiency is achieved for the order diffracted along the surface.
Actually, the efficient coupling of incident light with a BG depends on several factors including the incidence angle and the BG parameters (such as blaze angle and grating depth) and a theoretical assessment typically requires a numerical approach. Commercially available BGs are optimized for a specific wavelength by maximizing the reflection coefficient in the Littrow configuration, for which directions of the incident light and the order of diffraction overlap. 

By studying a set of BGs as targets, a very strong enhancement of the bunch charge, up to $~\approx 660$~pC per bunch, was found for a BG having a blaze angle of $13^{\circ}$ (Fig.\ref{fig:blazed}), optimized for $0.75~\mu$m wavelength according to the above mentioned criterion. The same grating also yielded the highest cut-off energy of electrons amongst the tested BGs, although the enhancement was less dramatic with respect to that of the bunch charge. 

It should be stressed that the investigation of BGs was limited to commercially available types and that so far no further optimization study or design of a specific BG has been completed. In addition, the available BGs had a metallic coatings which could favor the formation of a preplasma at the surface, affecting the interaction in an uncontrolled way. Nevertheless, the preliminary study of BGs reveal their potential for optimization of the SP coupling and secondary source performance, and is a further demonstration of the sensitivity of high-contrast femtosecond interaction to details of the surface structuring on a sub-micrometer scale. 

\section{High harmonic generation} 
\label{sec:harmonics}

\subsection{Experimental observation}

In the interaction of short intense laser pulse with solid targets, high harmonics (HHs) of the laser frequency are emitted (see reviews\cite{teubnerRMP09,thauryJPB10} and references therein). For the typical laser wavelength of $0.8~\mu$m, the highest harmonic order $m$ may be high enough that the frequency reaches the extreme ultraviolet (XUV) range, and the coherence of the HHs results in a temporal structure of a train of attosecond pulses. 

Whatever the details of the generation mechanism, from a flat surface the HHs are emitted in the specular direction, collinear to the reflected light at the first fundamental harmonic. Foreseen applications of the HHs may require their angular separation, which has been achieved using grating targets\cite{yeungOL11,yeungNJP13,cerchezPRL13,zhangOE17} so that each HH is emitted at angles $\phi_{mn}$ determined by the grating equation:
\begin{equation}
\frac{n\lambda}{md} = {\sin(\phi_{\rm i}) + \sin(\phi_{mn})}
\; ,
\label{eq:diffHH}
\end{equation}
where $n$ is the diffraction order and $\phi_{\rm i}$ is the angle of incidence. 
In our work we aimed to combine such angular separation with possible enhancement produced by SP excitation, when irradiating the grating at a resonant angle.  PIC simulations\cite{fedeliAPL17} performed to design the experiment showed that at the SP resonance both the highest value observed for $m$ and the associated intensity increased for HHs within angles compatible with those predicted by Eq.(\ref{eq:diffHH})), the enhancement being particularly strong for HHs emitted close to the tangent at the target surface.

While the preliminary PIC simulations were performed for a target with a step boundary profile, in the experiment a density profile with finite, sub-micron scalelength was produced by a controlled prepulse\cite{kahalyPRL13}. As previously demonstrated in flat targets, such tailoring of the density profile leads to order-of-magnitude increase in the HH efficiency, and in the conditions of the experiment performed at SLIC this was essential for the detection of the signal. It is remarkable that the scalelength of the prepulse-produced density profile was of the same order of the grating depth, but the periodic structure was not destroyed since the effects of HH diffraction and enhancement at the SP resonance were evident and in line with the theoretical expectations.

\begin{figure}[tb]
\includegraphics[width=0.48\textwidth]{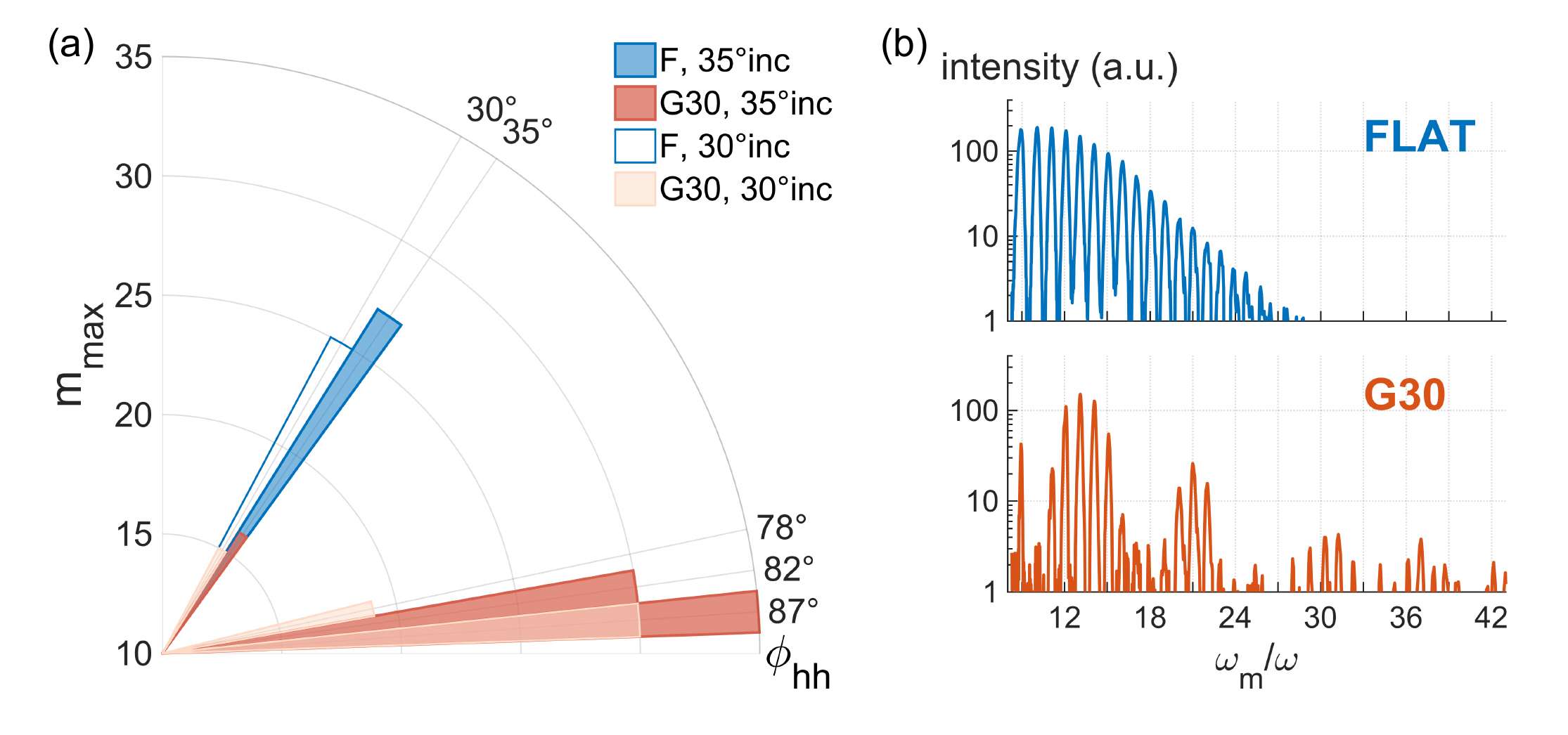}
\caption{Surface plasmon-enhanced high harmonic emission\cite{cantonoPRL18}. 
a): polar plot showing the maximum harmonic order observed $m_{\rm max}$ as a function of the observation angle $\phi_{\rm hh}$, for both a grating target (G30) with $\phi_{\rm res}=30^{\circ}$ at angles of incidence $\phi_{\rm i}=30^{\circ}$ and $35^{\circ}$, and a flat target at the same values of $\phi_{\rm i}$. b): comparison of harmonic spectra along $\phi_{\rm hh}=87^{\circ}$ for G30 and $\phi_{\rm hh}=35^{\circ}$ for the flat target. The grating depth was $0.25~\mu{m}\simeq 0.3\lambda$ and a femtosecond prepulse was used to create a preplasma with a typical extension $L \simeq 0.1-0.2\lambda$ from the original surface in optimal conditions for harmonic generation.
Adapted and partially reproduced with permission from Phys. Rev. Lett. \textbf{120}, 264803 (2018). Copyright 2018 American Physical Society.
\label{fig:harmonics-exp}}
\end{figure}

The experimental observations are summarized in Fig.\ref{fig:harmonics-exp}. The HH spectrum emitted at an angle $\phi_{\rm hh}=87^{\circ}$ is compared with the spectrum of the HHs emitted from flat targets in the specular direction. The comparison shows an increase of the highest detectable order from $m\simeq 27$ in the flat case to $m \simeq 40$ in the grating case. At lower values of $m$, the intensities of selected HHs are similar in the two cases, the spectrum from the grating being modulated as the result of the HH selection due to diffraction. Notice that the grating effects are stronger at an incidence angle of $\phi_{\rm i}=35^{\circ}$, slightly larger than the expected resonant value of $\phi_{\rm res}=30^{\circ}$. This is qualitatively consistent with the average density near the surface being lower than the solid value because of the preplasma production, so that the value of $\phi_{\rm res}$ increases in accordance with Eqs.(\ref{eq:SPdisp}--\ref{eq:SPres}).

Electron emission near the tangent direction was simultaneously measured with the HH signal. The optimal conditions for HH enhancement did not correspond to the highest energy and strongest collimation for electrons\cite{cantonoPRL18}, which was qualitatively in agreement with the dependence on the target density predicted by the simple model. However, the correlation between HH and electron emissions seems to play an important part in the HH generation mechanism supported by the SP, as indicated by the simulations discussed in the next section.

\subsection{Numerical simulation and generation mechanism}

\begin{figure*}[tb]
\includegraphics[width=\textwidth]{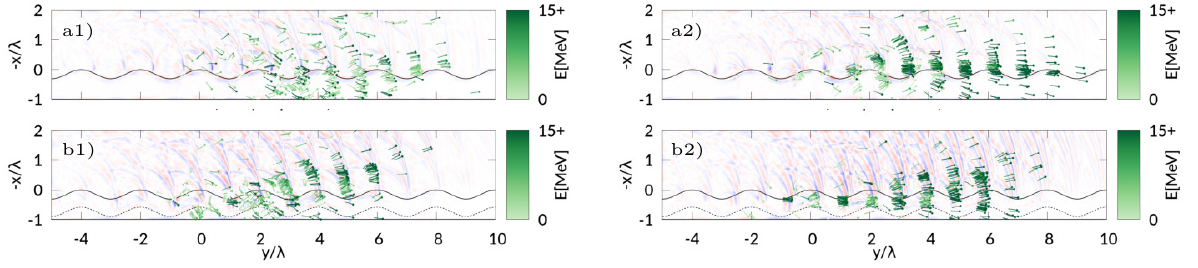}
\caption{PIC simulations of high harmonic generation in a grating\cite{cantonoPRL18}. Some electrons with energies above 10~MeV are represented by arrows parallel to their velocity and color coded (in green) according to their energy. The field $B_z$ is shown as a contourplot (in purple) after filtering out frequencies below the 7th harmonic of the incident laser. 
The line contour represents the density profile. 
a1), b1): snapshots from two simulations of a grating with a step-like density profile and $\phi_{\rm res}=30^{\circ}$, irradiated at $30^{\circ}$ and $35^{\circ}$ respectively, at the same instant $t=20\lambda/c$ from the start of the interaction.
a2), b2): same as a1) and b1) but for a grating with an exponential density profile to simulate the preformed density gradient. The line contours are the  $n_e=10n_c$ and $100n_c$ iso-density surfaces, respectively.   
\label{fig:harmonics-sim}}
\end{figure*}

PIC simulations in 2D reproduced the experimental observations and gave an insight of the particular HH generation mechanism in the presence of a SP. In the simulations, the electron density was $n_e=100n_c$, the target thickness was $1\lambda$, the laser pulse had a $\cos^2$ envelope and other target and laser parameters were the same as in the experiment (see Ref.\onlinecite{cantonoPRL18} for details).
Fig.\ref{fig:harmonics-sim} shows snapshots from the simulations where some high energy electrons have been indicated by arrows parallel to their velocity. In order to evidentiate HH radiation on the same plot, the magnetic field $B_z$ (perpendicular to the plane of the simulations) is shown after filtering out its low harmonics components. The simulations are performed both without and with the preformed density gradient, the HH intensity being stronger in the latter case, in agreement with the experimental observations. 

The HH pulses appear to be spatially correlated to the bunching of the high energy electrons in sub-wavelength structures, regularly spaced with a period close to the laser and SP wavelengths. More precisely, the HH pulses appear to propagate immediately ahead of the electron bunches. The scenario emerging from the simulations, in our current interpretation, is that electrons are accelerated by trapping in the SP potential wells, which also leads to bunching. 
The bunched electrons, being strongly relativistic, scatter the laser radiation preferentially along their velocity, resulting in collimated emission in a near-tangent direction. 

In principle, XUV radiation in a direction close to that of the accelerated electrons could also be produced via the Smith-Purcell effect\cite{smith-purcellPRL53}. The wavelength $\lambda_{\rm rad}$ observed at an angle $\phi$ from the target normal would be given by
\begin{equation}
  \lambda_{\rm rad}=\frac{d}{n}\left(\frac{1}{\beta_{\parallel}}-\cos\phi\right) \; ,
  \label{eq:SmithPurcell}
\end{equation}
where $\beta_{\parallel}$ is the component of the electron velocity (normalized to $c$ parallel to the grating plane; we assume the diffraction order $n=1$ in the following estimates. By taking $\sim 10$~MeV electrons moving parallel to the grating and $\phi=87^{\circ}$ as in our set-up, Eq.(\ref{eq:SmithPurcell}) gives $\lambda_{\rm rad} \simeq \lambda/185$ which falls outside both the range of our XUV spectrometer and the spatial resolution of our simulations. However, for strongly relativistic electrons and near-tangent radiation emission the result is strongly dependent upon the actual angle $\phi_e$ at which the electrons move with respect to the grating normal. In addition, both the energy and the angular spread of electrons would make the radiation broadband. Thus, while in principle Smith-Purcell radiation could contribute to the XUV emission, it cannot account for the harmonic structure observed in our experiments and simulations.\footnote{In Ref.\onlinecite{cantonoPoP18} the \emph{inverse} Smith-Purcell effect\cite{mizunoN75} was discussed and ruled out as a possible mechanism leading to electron acceleration along the grating.}

\section{Concept for quasi-single-cycle plasmons}
\label{eq:singlecycle}

The most peculiar characteristic of the secondary sources of energetic radiation (including ions, electrons and photons) driven by short pulse lasers is their duration, which may open new possibilities in ultrafast science. For example, if the electron bunch duration can be shortened down to the few fs, or even sub-fs range (which obviously needs a monoenergetic spectrum to avoid velocity dispersion), electron diffraction becomes able to resolve atomic and molecular motions\cite{zewailARP06,sciainiRPP11,millerS14}. HH generation from solids can provide a source for applications in attosecond science\cite{hentschelN01,baltuskaN03,agostiniRPP04,cavalieriN07,corkumNP07,krauszRMP09} if a single as pulse can be isolated from other pulses in the train. A strategy for such aim is the use of a laser pulse with wavefront rotation\cite{quereJPB14} (WFR), which was shown to separate attosecond pulses in different directions (``lighthouse effect'').

We are currently investigating the effects of WFR in the context of SP excitation on gratings, in order to generate shorter SP-enhanced HH and electron emissions. The basic idea is that when a pulse with induced WFR is incident on a surface, the rotation of the phase front is equivalent to a continous rotation in time of the angle of incidence. Hence, when such pulse impinges on a grating near the resonant angle for SP excitation, the resonant condition is satisfied for a time shorter than the laser pulse duration and determined by the angular velocity of WFR. In other words, WFR may provide temporal gating of the SP resonance.

\begin{figure}[b!]
\includegraphics[width=0.49\textwidth]{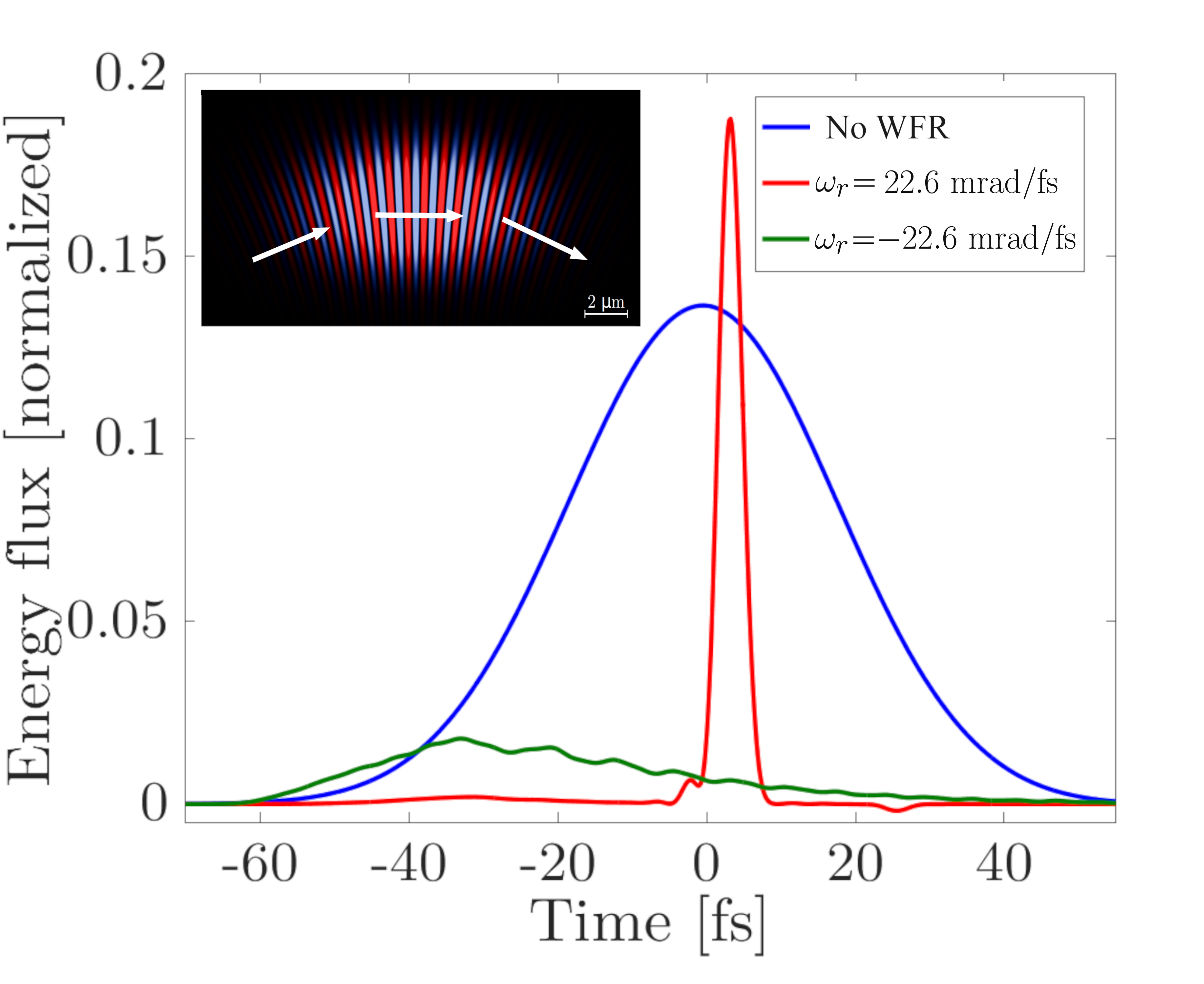}
\caption{
Generation of a near-single cycle surface plasmon by a laser pulse ($\lambda=0.8~\mu$m) with wavefront rotation\cite{pisaniACSP18} in simulations with the Meep code. The blue, red and green line show the energy flux of plasmons excited by a laser pulse of 29.5~fs duration with zero, positive and negative rotation, respectively. The angular velocity of wavefront rotation $\omega_r$ corresponds to 0.06~rad per cycle. The inset shows the 2D field profile of the pulse at waist.
Adapted and partially reproduced with permission from ACS Photonics \textbf{5}, 1068 (2018). Copyright 2018 American Chemical Society.
\label{fig:WFR}
}
\end{figure}

So far we have tested the concept in the linear regime, simulating the interaction of a WFR pulse with a metallic grating\cite{pisaniACSP18} using the Meep FDTD code\cite{oskooiCPC10}. Fig.\ref{fig:WFR} shows results obtained for a laser pulse of $29.5$~fs duration and $0.8~\mu$m wavelength impinging on a grating at the resonant angle of $25^{\circ}$. For a WFR angular velocity $\omega_r=22.6~\mbox{mrad fs}^{-1}$, i.e. 0.06~rad/cycle, the SP has a duration of $\sim$3.8~fs, which corresponds to $\sim$1.4 laser cycles. For such extreme durations, carrier-envelope or absolute phase effects become relevant in characterizing the temporal profile of the SP field. 

\begin{figure}[t!]
\includegraphics[width=0.49\textwidth]{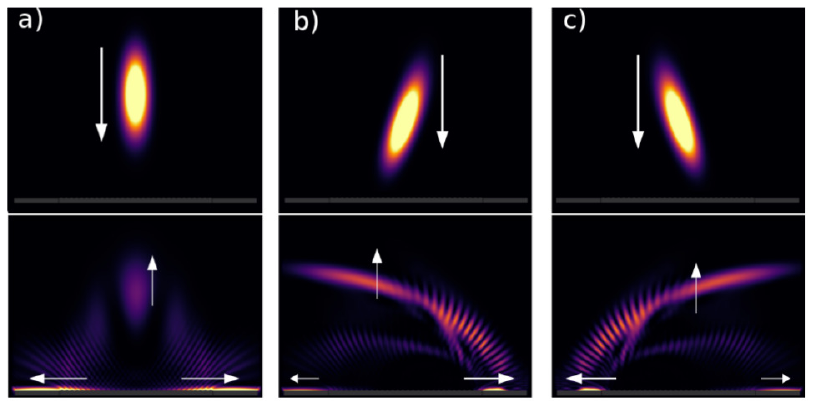}
\caption{Effect of the sign of wavefront rotation on surface plasmon generation.  Two snapshots from Meep simulations at normal laser incidence on a grating are shown for a) no rotation, b) positive rotation, c) negative rotation. While in case a) two symmetrical plasmons are generated, in cases b) and c) one plasmon is shorter and stronger than the other one. The parabolic-shaped signal is due to the scattering from the plasmon propagating along the grating. 
\label{fig:WFRsign}
}
\end{figure}

As also apparent in  Fig.\ref{fig:WFR}, the sign of $\omega_r$ is crucial to achieve the shortening effect. Based on our observations in the Meep simulations, we define the positive rotation as follows. Because of the WFR, each phase front impinges at a different angle on the target surface, which implies that the midpoint of each front (where the field amplitude has a maximum) slides along the surface. Thus, the centroid of the intensity distribution moves along the waist plane with a velocity that may be estimated as $v_{\rm i} \sim \lambda\omega_r\cos\phi_{\rm i}$, the sign being dependent on that of $\omega_r$. The value of $v_{\rm i}$ increases when displacing the waist from the target plane, which can be used as an additional parameter to optimize the coupling. 
We found that $v_{\rm i}$ must be in the same direction as the SP propagation velocity in order to achieve the shortening effect, which also comes with the peak amplitude of the few-cycle SP being higher than that of the ``ordinary'' SP excited without WFR. In the case of negative rotation, the excited SP is much weaker and not significantly shorter than the ordinary SP.

The effect of the sign of $\omega_r$ may be evidenced by comparing simulations in which the laser pulse impinges at normal incidence ($\phi_{\rm res}=0$) on a grating with period $d=\lambda$. In this particular case, two oppositely propagating SPs are excited. Fig.\ref{fig:WFRsign} shows that without WFR the two SPs are symmetrical. For positive $\omega_r$ (corresponding to the intensity contours of the laser pulse being inclined towards the right direction in Fig.\ref{fig:WFRsign}), the SP propagating from left to right is much shorter and stronger than the SP in the opposite direction. The situation is reversed for negative $\omega_r$. The different amplitude of the plasmons can be also noticed in the different intensity of the field scattered by the plasmon (the parabolic tail structure in the bottom images of Fig.\ref{fig:WFRsign}) while propagates along the grating. Such scattering can be considered as the inverse effect of plasmon excitation by an EM wave impinging on a grating.

\section{Conclusions and outlook}

The experiments reviewed in this paper demonstrated that using high contrast femtosecond pulses it is possible to excite propagating surface plasmons on grating targets at relativistically strong intensities. These findings open up new possibilities for applications inspired by either  ``ordinary'' (low field) plasmonics or high intensity laser-plasma physics. For example, the acceleration of electrons by surface plasmons can provide, with further optimization and characterization, a useful source of multi-MeV electrons for applications requiring high charge fluxes. The generation of high harmonic XUV pulses with sub-femtosecond duration and quasi collinear with the electron beam may also be of interest for pump-probe applications. 

On this route to surface plasmon-based sources development and optimization, our most recent experiments have shown that it is possible to exploit the tailoring of the density profile on a sub-micrometer scale, either as pre-imposed (blazed gratings) or dynamically generated, i.e. optically induced ``preplasmas''. The latter finding suggests that it may be even possible to exploit optical generation of the grating structures\cite{monchocePRL14,leblancNP17}. 
Moreover, our proposed concept for the temporal gating of the surface plasmon resonance in order to generate single-cycle plasmons may find application in ultrafast plasmonics, also at low fields. On the high field side, developing the potential of relativistic surface plasmons will also benefit from a deeper theoretical understanding of their nonlinear behavior. 

In conclusion, our investigations are a piece of work characterized by two words, \emph{femtosecond} and \emph{nanometer}, amongst the four words that ``capture the essence of solid state plasmas and plasmonics'' according to Manfredi\cite{manfrediPoP18}. As a general perspective, the control and exploitation of SPs on the nanoscale in space and femtoscale in time are at the frontier of plasmonics and strongly related to nonlinear effects\cite{stockmanOE11,kauranenNphot12}. 
While our results are just a very first step towards pushing the frontier towards the regime of superintense fields and sub-femtosecond duration, we expect that this extreme nanoplasmonics may further emerge as an active and promising research area across the traditional borders of plasmonics, attophysics, and relativistic plasma physics.

\begin{acknowledgments}
We are grateful to C.~Riconda and A.~Sgattoni for their longstanding collaboration on the topic of high field plasmonics, and to F.~Amiranoff for enlightening discussions. We also acknowledge the contributions of L.~Chopineau, A.~Denoeud, D.~Garzella, F.~Reau, I.~Prencipe, M.~Passoni, M.~Raynaud, M.~Kv\v{e}to\v{n}, and J.~Proska to the research reviewed in this paper, 
the help by J.-P.~Larbre (ELYSE, Universit\'e Paris Sud, Orsay, France) in the calibration of the Lanex screen,
and the overall support of the Saclay Laser Interaction Center team.
Simulations were performed using the PICcante code at the HPC Cluster CNAF (Bologna, Italy), with the precious help of S.~Sinigardi.
This research received financial support from 
LASERLAB-EUROPE (Grant agreement No. 284464, EU FP7),
Investissement d'Avenir LabEx PALM (Grant ANR-10-LABX-0039), 
Triangle de la physique (Contract No. 2014-0601T),
and 
Agence Nationale pour la Recherche (Grant No.ANR-14-CE32-0011).
G.~C. acknowledges partial support by the Universit\'{e} Franco-Italienne (Vinci program 2015, Grant No. C2-92).
L.~F. acknowledges partial support by the European Research Council Consolidator Grant ENSURE (ERC-2014-CoG No. 647554).
\end{acknowledgments}

%



\end{document}